\documentclass{elsart}
\usepackage[dvips]{graphicx}
\usepackage{bm}
\usepackage{psfrag}
\usepackage{amsmath}
\usepackage{mathrsfs}
\usepackage{amssymb}
\usepackage{cite,mcite}
\newcommand{\beq}{\begin{equation}}
\newcommand{\eeq}{\end{equation}}
\newcommand{\bea}{\begin{eqnarray}}
\newcommand{\eea}{\end{eqnarray}}
\newcommand{\nd}{\noindent}
\newcommand{\pola}{\frac{1}{2}}
\newcommand{\omc}{\Omega_c^0}
\newcommand{\omm}{\Omega^-}
\newcommand{\xic}{\Xi_c^+}

\begin{document}


\begin{frontmatter}

\title{Cabibbo-suppressed decays of the $\omc$ --- feedback 
to the $\xic$ lifetime}

\author{A. Babi\' c,}
\ead{ababic@thphys.irb.hr}
\author{B. Guberina,}
\ead{guberina@thphys.irb.hr}
\author{B. Meli\' c,}
\ead{melic@hippo.irb.hr}
\author{H. \v Stefan\v ci\' c}
\ead{shrvoje@thphys.irb.hr}
\address{Theoretical Physics Division, Rudjer Bo\v skovi\' c 
Institute, P.O. Box 180,HR-10002, Zagreb, Croatia}

\begin{abstract}
We investigate a possible background of the type $\omc \rightarrow 
\xic \pi^-$ to the CLEO $\xic$ lifetime measurement. This decay mode 
may lead to an overestimate of the $\xic$ decay length and, therefore, 
increase the measured $\xic$ lifetime. The branching ratio $\Gamma 
(\omc \rightarrow \xic \pi^- )/\Gamma (\omc \rightarrow \Omega^- 
\pi^+ )$ is analyzed in the framework of the pole model and the 
modified current algebra. We find that the $\omc \rightarrow \xic 
\pi^- $ decay mode could not generate a substantial systematic error 
in the $\xic$ lifetime measurement. Also, it cannot significantly 
reduce the disagreement between theoretical and experimental values 
of the $\xic$ lifetime.
\end{abstract}

\end{frontmatter}

\thispagestyle{empty}

The lifetime measurements of charmed baryons are well known
\cite{Bellini:1997ra,*Mannel:2001ux} to be very important in 
estimating and disentangling the different preasymptotic effects in 
the decays of charmed hadrons. They provide the most direct way to 
determine the weak mixing angles and to test the unitarity of the CKM 
matrix. The CP violation outside the kaon system can be studied, and 
one can also test our present knowledge of the QCD confinement inside 
hadrons. 

The preasymptotic effects \cite{Blok:1991st}, like the inclusion 
of soft degrees of freedom (light quarks, gluons) generate 
nonperturbative power corrections, e.g. the destructive and/or 
constructive Pauli interference, and the W-exchange contribution, 
producing the diversity of lifetimes of charmed mesons 
\cite{Guberina:1979xw,Bilic:1984nq,Shifman:1985wx} and baryons 
\cite{Shifman:1986mx,Guberina:1986gd}, which would, otherwise, 
be all equal in the asymptotic limit of infinitely heavy quark 
mass\footnote{It appears astonishig that decay rates of weak and 
radiative decays are described in terms of few basic quantities, 
e.g. quark masses, and hadronic expectation values of several 
leading local operators.}. 

Inclusive hadronic decay rates and lifetimes were calculated a 
long time ago \cite{Guberina:1979xw,Bilic:1984nq,Shifman:1985wx,%
Shifman:1986mx,Guberina:1986gd} by summing over all possible channels 
and integrating over some range of energies. A `practical' version of 
the OPE is used in calculations, i.e. it is assumed that the 
coefficient functions can be found perturbatively and all 
nonperturbative effects reside in matrix elements. In real world, 
however, there are nonperturbative effects even at short distances, 
and the matrix elements are subject to perturbative corrections too. 

Suprisingly enough, the theory works rather well, even in the charmed
hadron sector, although the expansion parameter $\sqrt{\mu_G^2(D)/
m_c^2}\simeq 0.5$ is not really small (the corresponding parameter in 
beauty decays is $\sqrt{\mu_G^2(B)/m_b^2}\simeq 0.13$).

A systematic study of charmed baryon decays was performed a few years 
ago \cite{Guberina:1998yx}, with good agreement between theory and 
experiment. The theoretical predictions were rather stable to the 
uncertainties in the wave functions of heavy baryons and/or to the 
choice of the renormalization/factorization scale, except in the case 
of the $\xic$ charmed baryon. It was not clear if the peculiar 
behavior of the $\xic$ was a pure coincidence due to the wild 
cancellation of different preasymptotic effects, or some deeper 
understanding was missing. The theoretical result, $\tau (\xic )_{
\mathrm{th}}=0.27$ ps, for $m_c=1.35$ GeV, $\Lambda_{\mathrm{QCD}}=
300$ MeV, had to be compared with the experimental value, $\tau (\xic 
)_{\mathrm{exp}}=(0.35 \pm 0.07)$ ps. The difference, at that time, 
was not so significant that one would have had to worry. However, it 
was clear that future more precise measurements could disturb an 
idyllic concordance between theory and experiment.

\begin{figure}[t]
\centering
\psfrag{a}{WA62${}_{85}$}
\psfrag{b}{E400${}_{87}$}
\psfrag{c}{Accmor${}_{89}$}
\psfrag{d}{E687${}_{93}$}
\psfrag{e}{E687${}_{98}$}
\psfrag{f}{FOCUS${}_{01}$}
\psfrag{g}{CLEO${}_{01}$}
\psfrag{h}{PDG${}_{00}$}
\psfrag{i}{PDG${}_{02}$}
\psfrag{x}[ct]{Lifetimes (ps)}
\includegraphics[scale=0.45]{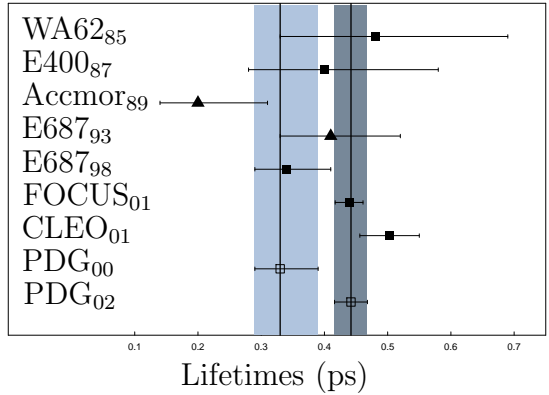}
\caption{ $\xic$ lifetime experiments. The left (right) 
band is the 1$\sigma$ PDG 2000 \protect\cite{Groom:2000in}
(2002 \protect\cite{Hagiwara:2002fs}) world average. E687
$_{93}$ is excluded from the PDG 2000 value and Accmor
from the PDG 2002 value.}
\label{experimenti}
\end{figure}

Fig. \ref{experimenti} shows the results of $\xic$ lifetime 
expertiments performed up to now. One can see that two new 
measurements with significantly improved accuracy, FOCUS 
\cite{Link:2001qy} and CLEO \cite{Mahmood:2001em}, are above the 
previous world average $1\sigma$ margin, in the case of CLEO, even
above the $2\sigma$ margin. By including these two new measurements 
the average has changed from 0.33 ps to 0.442 ps. In particular, 
FOCUS precisely measured the charmed-strange baryon $\xic$ lifetime 
as
\beq
\tau (\xic)=0.439 \pm 0.022 \pm 0.009 \text{ ps}.
\eeq
In the FOCUS spectrometer, which is well suited to reconstruct 
short-lived charmed decays, the charmed particles are produced as 
the product of the interaction between high energy-photons with 
$\langle E \rangle \simeq 180$ GeV in a segmented BeO target and an 
excellent vertex separation between the production and decay vertices 
is provided by two silicon vertex detectors.

All previous experiments, including that performed by FOCUS, are 
fixed-target experiments. CLEO performed the only colliding beam 
experiment. Therefore, it has different systematics and different 
backgrounds. In spite of the fact that the charmed baryon lifetimes 
are not measured as precisely as those of charmed mesons, CLEO and 
SELEX \cite{Bianco:2003vb} recently measured $\tau (\Lambda_c^+)$ to a 
precision of 5\%. Other charmed baryons ($\xic$, $\Xi_c^0$, $\Omega_c$) 
are measured with up to an uncertainty of 30\%. CLEO's measurement 
gives
\beq
\tau(\xic)=0.503\pm 0.047 \text{ (stat.)} 
\pm 0.018 \text{ (syst.) ps.}
\eeq
This result is obtained using an integrated luminosity of $9.0\text{ 
fb}^{-1}$ of $e^+ e^-$ annihilation data taken with the CLEO IV.V 
detector at the CESR. The data were taken at energies at and below 
the $\Upsilon$(4S) resonance and include $\sim 11\cdot 10^6$ $e^+ e^- 
\rightarrow c {\bar{c}}$ events. The $\xic$ is reconstructed from the 
$\Xi^- \pi^+ \pi^-$ decay mode. Each $\Xi^-$ is reconstructed from 
$p \pi^-$. The assumption is that the $\xic$ is produced at the primary 
event vertex and is not a decay product of another weakly decaying 
particle, e. g. $\Omega_c^0 \rightarrow \xic \pi^-$, $\Xi_{cc}^{++}
\rightarrow \xic + \cdots$.

\begin{figure}[b]
\centering
\psfrag{a}[lb]{\parbox{3cm}{\scriptsize 
\centering $\xic$ production \\vertex}}
\psfrag{b}[cb]{\parbox{2cm}{\scriptsize 
\centering $\xic$  decay \\vertex}}
\psfrag{d}[cb]{\parbox{2cm}{\scriptsize 
\centering PE / $\omc$ \\ production \\vertex}}
\psfrag{f}{$\Delta$}
\psfrag{c}[cb]{\scriptsize $y_{\mathrm{decay}}- 
y_{\mathrm{production}}$}
\includegraphics[width=0.7\linewidth,height=0.35\linewidth]{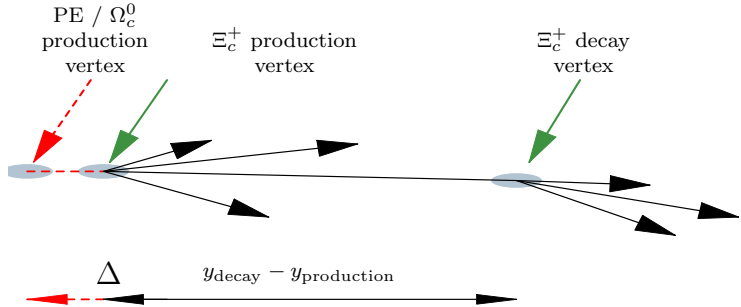}
\caption{$\omc $ is produced at the primary 
event (PE) vertex and decays into $\xic$.}
\label{event1}
\end{figure}

If $\omc$ is produced at the primary event (PE) vertex, travels a 
certain distance and decays into $\xic$ and $\pi^-$ (Fig. \ref{event1}), 
the production vertex  of $\xic$ is misinterpreted to be at the PE 
vertex and there is an addition ($\Delta$) to the measured proper time:
\beq
t=\frac{m_{\xic}}{c p_{y_{\xic}}}
(y_{\mathrm{decay}}-y_{
\mathrm{production}} +\Delta ).
\eeq
\nd The measured $\xic$ lifetime will be shifted towards a 
higher value. 

The purpose of this letter is to examinate the relevance of the $\omc
\rightarrow\xic\pi^-$ decay mode as a possible source of a systematic 
error for the $\xic$ lifetime measurement. To this end, we study the 
ratio of two exclusive decay modes,
\beq
\eta =\frac{\Gamma(\omc\rightarrow\xic 
\pi^-)}{\Gamma (\omc\rightarrow\omm\pi^+)}		\label{eq:eta}
\eeq
\nd for the following reasons: the $\omc\rightarrow \omm \pi^+$ 
process is expected to be one of the first and best measured $\omc$
exclusive decays in the near future; therefore it is quite convenient 
to have the contribution of $\omc\rightarrow\xic \pi^-$ normalized to 
the rate of $\omc\rightarrow\omm\pi^+$ \cite{Kass}; the $\omc
\rightarrow\omm\pi^+$ process has a factorizable contribution only, 
which reduces theoretical uncertainties; uncertainties are further 
suppressed by considering ratios of exclusive decay widths.


\paragraph*{$\bm {\omc \rightarrow \xic \pi^-}$ decay mode} In the 
$\omc \rightarrow \xic \pi^- $ decay, the decay happens in the 
light-quark sector and the pion emerges with a momentum of $O$(200 MeV) 
that can be considered 'reasonably' soft. Therefore, there is a 
similarity between this decay and the hyperon ($\Delta S=1$) decays 
for which the soft-pion limit technique with pole corrections was 
successfully applied \cite{LeYaouanc:1979ef,*Borasoy:1999ip} with the 
predictions for the branching fractions within 20\% from experimental 
values. We believe that the soft-pion limit is equally applicable to the 
$\omc \rightarrow \xic \pi^-$  decay. 

The invariant amplitude for the decay of the initial baryon $B_i 
(1/2^+)$ to the final baryon $B_f (1/2^+)$ and a pion $\pi^a$, 
$a=1,2,3$, is given by
\beq
\langle B_f \pi^a | \mathscr{H}_{\mathrm{W}}
(0) |B_i\rangle = \mathrm{i} \, \overline{u}_f 
(A-B \gamma_5 )u_i,					\label{invamp}
\eeq
\nd with A and B to be determined using the standard nonleptonic weak
hamiltonian
\beq
\mathscr{H}_{\mathrm{W}}=\sqrt{2} G_F 
V_{\bar{q}_3 \, q_4}V_{\bar{q}_1 \, q_2}^*
(c_- O_- +c_+ O_+),
\eeq
\nd where $O_{\pm}$ are local 4-quark operators
\beq
O_{\pm}=(\bar{q}_{1L}\gamma_{\mu}q_{2L})
(\bar{q}_{3L}\gamma^{\mu}q_{4L})\pm (
\bar{q}_{3L}\gamma_{\mu}q_{2L})(
\bar{q}_{1L}\gamma^{\mu}q_{4L}),
\eeq
\nd with 
$(\bar{q}_{iL}\gamma_{\mu}q_{jL})=\pola\bar{q}_i\gamma_{\mu}
(1-\gamma_5)q_j$, and $V$'s are the elements of the CKM matrix. The 
Wilson coefficients in the leading logarithmic approximation are 
given by 
\beq
c_{\pm}(\mu^2)\cong\left(\frac{\alpha_s(
\mu^2)}{\alpha_s(M_{\mathrm{W}}^2)}
\right)^{d_{\pm}/2b},
\eeq
\nd where $b=\frac{1}{3}(11 N_{\mathrm{c}}-2 n_{\mathrm{f}})$, 
$N_{\mathrm{c}}$ and $n_{\mathrm{f}}$ being the number of colors 
and flavors, respectively. The quantities $d_-=-2d_+=8$ are 
proportional to the anomalous dimensions of the operators 
$O_-$ and $O_+$.

In the approach of references \cite{LeYaouanc:1979ef,Stech:1991fx}  
modified current algebra techniques were applied, i.e. the soft-pion 
amplitude (commutator term) was corrected for the soft-pion limit. 
The contribution coming from baryon poles is given as 
\bea
A^{\mathrm{CA}}&=&A^{\mathrm{soft}}+A^{
\mathrm{corr}}\nonumber\\
&=&\frac{\sqrt{2}}{f_{\pi}}\langle 
B_f|\left[ Q^a,\mathscr{H}_{\mathrm{W}}^{
\mathrm{PC}}\right]|B_i\rangle\nonumber\\
&& -\frac{\sqrt{2}}{f_{\pi}}
\sum_{B_n^*\left(\frac{1}{2}^-\right)}
(m_{B_f} -m_{B_i})\left( \frac{g_A^{B_f 
B_n^* }b_{B_{n}^* B_{i}}}{m_{B_i} -
m_{B_n}^*}+\frac{b_{B_{f}B_{n}^*}g_A^{ 
B_n^* B_i}}{m_{B_f} -m_{B_n}^*}\right),\\
B^{\mathrm{pole}}&=&\!\!\!\!\sum_{B_n 
(\frac{1}{2}^+)}\!\!\!\frac{\sqrt{2}}{f{\pi}}
\left(\frac{m_{B_f} +m_{B_n}}{m_{B_i}-
m_{B_n}} g_A^{B_f B_n }a_{B_{n} B_{i}}+
\frac{m_{B_i} +m_{B_n}}{m_{B_f} - m_{B_n}}
a_{B_{f}B_{n}}g_A^{ B_n B_i}\right) . 			\label{eq:AB}
\eea
\nd In the above equations the S- and P-wave amplitudes are calculated 
in the framework of the pole model. Using the Lehman-Symanzik-Zimmerman 
reduction formalism, the pion momentum is taken off shell. The pion 
field is related to the axial vector curent via PCAC, and a complete 
set of states is inserted.

In (\ref{eq:AB}), the  baryon-baryon weak matrix elements $b_{B_{j}
B_{i}}^*$ and $a_{B_{j}B_{i}}$  are defined as
\beq
\langle B^*_j ( 1/2^-)|\mathscr{H}_{
\mathrm{W}}^{\mathrm{PV}} | B_i\rangle 
= \mathrm{i} \, b_{B_j B_i}^*{\bar{u}}_j u_i \, , 	\label{wme1}
\eeq
\beq 
\langle B_j ( 1/2^+)|\mathscr{H}_{
\mathrm{W}}^{\mathrm{PC}} | B_i\rangle 
= a_{B_j B_i}{\bar{u}}_j \gamma_5 u_i \, ,		\label{wme2}
\eeq
\nd and $g_A^{B_{i}B_{j}}$ is the axial form-factor, related to the 
strong $g^{B_{i}B_{j}M}$ baryon-baryon-meson coupling through the 
generalized Goldberger-Treiman relation. The pion decay constant 
$f_{\pi}$ is taken as $0.132$ GeV. The weak matrix elements 
(\ref{wme1}) and (\ref{wme2}) and the axial form-factors are 
calculated inside the MIT bag model.

\begin{figure}
\psfrag{a}[c]{$s$}
\psfrag{b}[c]{$s$}
\psfrag{c}[c]{$c$}
\psfrag{d}[c]{$s$}
\psfrag{e}[c]{$c$}
\psfrag{f}[c]{$u$}
\psfrag{g}[c]{${\bar u}$}
\psfrag{h}[c]{$d$}
\psfrag{i}[cB]{$s$}
\psfrag{j}[cB]{$c$}
\psfrag{k}[cB]{$d$}
\psfrag{l}[cb]{$[\Xi_c^{0}]$}
\includegraphics[width=7cm]{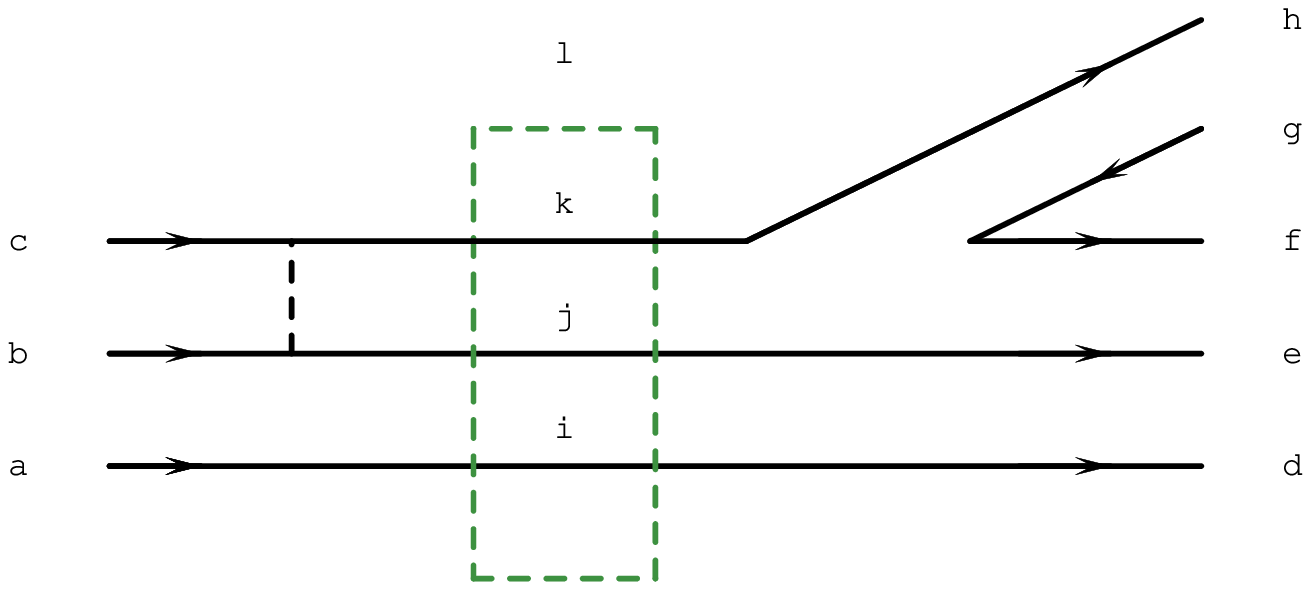}
\psfrag{a}[c]{$\omc$}
\psfrag{b}[lB]{$\xic$}
\psfrag{c}[cb]{$a_{\Xi_c^{'0}\omc}$ 
($b_{\Xi_c^{*0}\omc}$)}
\psfrag{d}[c]{$\Xi_c^{'0}(\Xi_c^{*0})$}
\psfrag{e}[lB]{$\pi^-$}
\psfrag{f}[cb]{$g^{\xic \Xi_c^{'0}\pi}
(g^{\xic \Xi_c^{*0}\pi})$}
\psfrag{g}{}
\psfrag{h}[c]{}
\includegraphics[width=7cm]{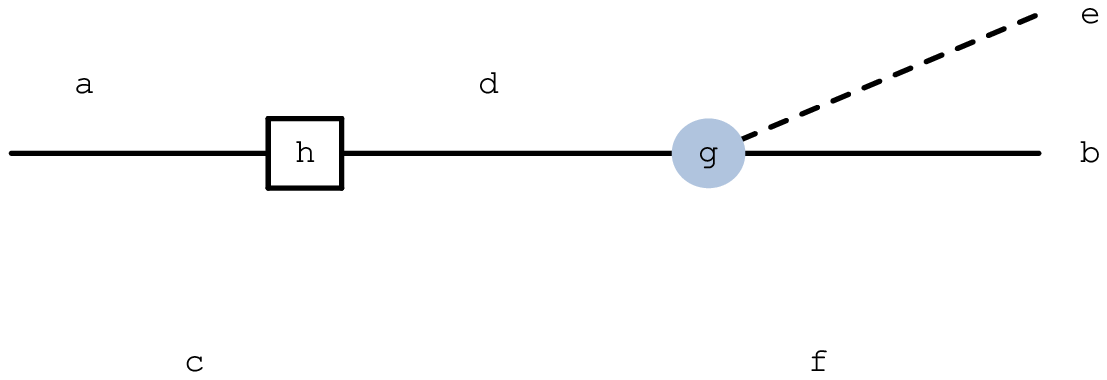}
\caption{Pole diagrams for the $\omc \rightarrow 
\xic \pi^-$ decay mode: at the quark level and 
in terms of effective couplings.}
\label{slika1}
\end{figure}

Concerning the pole resonances, the only flavor structure that can 
be formed in an intermediate state of the $\omc \rightarrow \xic 
\pi^-$ decay is ($dsc$), (Fig.\ref{slika1}). The main contribution to 
the S-wave amplitude comes from the commutator term in (\ref{eq17}), 
providing a simple means of summing contributions from all 
intermediate states in the soft-pion limit. The correction to this 
term is dominated by ($1/2^-$) resonances, the lowest one being for 
our decay $\Xi_c^{0}(2790)$ (denoted by $\Xi_c^{*0}$). P-wave 
amplitudes are dominated by the lowest lying ($1/2^+$) baryon 
intermediate states. Since the charmed antitriplet - antitriplet 
axial form factors  vanish, $g_A^{B_i(\bar 3) B_j(\bar 3)}=0$ , the 
lowest lying $\Xi_c^{0}$ resonance belonging to the charmed baryon 
antitriplet does not contribute. The main contribution to the P-wave 
amplitude comes from the $\Xi_c^{'0}$ baryon ($1/2^+$) state, 
belonging to the charmed baryon sextet. Therefore we have 
\beq
A^{\mathrm{CA}}=\frac{1}{f_{\pi}}
\langle \Xi_c^0|\mathscr{H}_{\mathrm{W}
}^{\mathrm{PC}}|\omc \rangle +\frac{1
}{f_{\pi}}\frac{m_{\omc}-m_{\xic}}{m_{
\omc} -m_{\Xi_c^{0*}}}g_A^{\xic \Xi_c^{*0} 
}b_{\Xi_c^{*0} \omc},
\eeq
\beq
B^{\mathrm{pole}}=\frac{1}{f_{\pi}}
\frac{m_{\Xi_c^{'0}} + m_{\Xi_c^{+}}}{
m_{\omc} - m_{\Xi_c^{'0}}}g_A^{\xic 
\Xi_c^{'0} }a_{\Xi_c^{'0} \omc}.			\label{eq17}
\eeq
\nd There is also a factorizable P-wave part of the  $\omc 
\rightarrow \xic \pi^-$ amplitude, which can be expressed as
\beq
B^{\mathrm{fact}} = -\frac{G_F}{\sqrt{2}}
V_{us}V_{ud}a_1 f_{\pi} (m_{\omc}+m_{
\xic})g_A^{\xic\omc},
\eeq
\nd where $a_1=\frac{1}{3}(2c_+ +c_-)$. The decay rate for $\omc 
\rightarrow \xic \pi^-$ is then given by 
\beq
\Gamma (\omc \rightarrow \xic + \pi^-)=
\frac{|\vec{p}_{\xic}|}{4\pi m_{\omc}}
\left[|A|^2(E_{\xic}+m_{\xic})+|B|^2 
(E_{\xic}-m_{\xic})\right].				\label{eq:gam1}
\eeq


\paragraph*{$\bm{ \omc \rightarrow \Omega^- \pi^+}$ decay mode.} This 
mode is of the type $B_f (1/2^+)\rightarrow B_i (3/2^+) + \pi$ and its 
invariant amplitude is 
\beq
\mathscr{M}=\mathrm{i} \, q_{\mu} \bar{u
}_f^{\mu} (B'-C\gamma_5 )u_i.
\eeq
\nd The expression for the decay rate is  
\beq
\Gamma (\omc \rightarrow \Omega^- + 
\pi^+)=\frac{|\vec{p}_{\xic}|^3 m_{\omc}}{
6\pi m^2_{\xic}}\left[ |B'|^2 (E_{\xic}+
m_{\xic})+|C|^2 (E_{\xic}-m_{\xic})\right] .		\label{width2}
\eeq
The $\omc \rightarrow \Omega^- \pi^+$ decay does not receive any 
pole contributions. There is only a factorizable P-wave amplitude 
contributing. This decay mode has already been calculated in the 
literature \cite{Xu:1992sw,Korner:1992wi,Cheng:1997cs} by applying 
different quark models. We have recalculated it in the MIT bag 
model in order to have a consistent calculation of the ratio $\eta$ 
(\ref{eq:eta}).


\paragraph*{Numerical results and discussions.} As we have already 
stated before, all our form factors, decay constants and matrix 
elements have been calculated in the MIT bag model. The calculations 
have been performed using the following parameter set: $\mu =1$ GeV, 
$\Lambda_{\mathrm{QCD}}=200$ MeV, particle masses are taken to be PDG 
average values \cite{Hagiwara:2002fs}, and MIT bag model parameters 
have the same values as in \cite{Cheng:1992sn}.

For the $\omc \rightarrow \xic \pi^- $ decay mode, we have the 
S-wave amplitude which is given by the current algebra term and a 
pole correction of 10\%. The P-wave amplitude has a factorizable 
contribution and a large pole contribution. Note from (\ref{eq:gam1}) 
that the P-wave amplitude is suppressed by a small kinematical 
factor, making contributions from S- and P-wave amplitudes of the 
same order of magnitude. The results of the calculation are sumarized 
in Table \ref{tabxi}. 
\begin{table}
\centering
\caption{Amplitudes ($\times 10^{7}$) and 
width (s$^{-1}$) for the $\omc \rightarrow 
\xic \pi^-$ decay mode. The invariant 
amplitude for the decay mode with the 
spin-$1/2$ particle in the final state is 
$\mathscr{M}=\mathrm{i} \, {\bar u}_f(A-B
\gamma_5)u_i$, with dimensionless S- and 
P-wave amplitudes.}
\label{tabxi}
\begin{tabular}{|cccc|ccc|c|}\hline
$A^{\mathrm{fact}}$  &  $A^{\mathrm{soft}}$  & 
$A^{\mathrm{corr}}$  &  $A^{\mathrm{tot}}$   & 
$B^{\mathrm{fact}}$  &  $B^{\mathrm{pole}}$  & 
$B^{\mathrm{tot}}$   &  $\Gamma$ (s$^{-1}$)\\ \hline
0       &   2.87   & 
0.25    &   3.12   & 
7.47    & -45.16   & 
-37.69  &   4.50$\cdot 10^{9}$ \\\hline
\end{tabular}
\end{table}
In the $\omc \rightarrow \Omega^- \pi^+$ decay the only 
nonvanishing contribution is from the factorizable part of 
the P-wave (B') amplitude, the D-wave (C) amplitude is zero 
(Table \ref{tabomega}). 
\begin{table}
\centering
\caption{Amplitudes ($\times 10^{7}$) and 
widths (s$^{-1}$) for the $\omc \rightarrow 
\Omega^- \pi^+ $ decay mode. The invariant 
amplitude for the decay mode with the 
spin-$3/2$ particle in the final state is 
$\mathscr{M}=\mathrm{i} \, q_{\mu}{\bar u
}_f^{\mu}(B'-C\gamma_5)u_i$, with P- and 
D-wave amplitudes having units GeV$^{-1}$.}
\label{tabomega}
\begin{tabular}{|c|c|c|}\hline
$B'$ (GeV$^{-1}$)  &  $C$ (GeV$^{-1})$  & 
$\Gamma$ (s$^{-1}$)\\ \hline
13.75  &  0  &  2.89$\cdot 10^{11}$ \\\hline
\end{tabular}
\end{table}
Finally, the ratio of partial decay rates two exclusive decay modes 
of $\Omega_c^0$ considered in this letter is 
\beq
\eta =\frac{\Gamma (\omc \rightarrow \xic 
\pi^- )}{\Gamma (\omc \rightarrow \Omega^-
\pi^+ )} = \frac{2.96\cdot 10^{-15}\text{ 
GeV}}{1.90\cdot 10^{-13}\text{ GeV}}=
0.016 \, .					\label{rezultat}
\eeq
The uncertainties, of order 10\%, are connected with the scale $\mu$
at which the Wilson coefficients are evaluated, whereas the variation
of $\Lambda_{\mathrm{QCD}}$ from 200 MeV to 300 MeV leads to  15\%
larger value of $\eta$.

The ratio of partial decay rates (\ref{rezultat}) shows that the 
branching ratios for the Cabibbo-suppressed decays of $\omc$ are 
at most at the level of a percent. The apparent dilatation of the 
$\xic$ baryon path due to the described cascade of weak decays 
from the initially formed $\omc$ baryon is quite small and 
certainly insufficient to explain the discrepancy of the recent 
$\xic$ lifetime measurements \cite{Link:2001qy,Mahmood:2001em} 
and theoretical calculations \cite{Guberina:1998yx}. This result 
is altogether not so surprising, although reassuring given that 
in exclusive decays there is always a possibility of a large pole 
contribution. 

Finally, it is worth mentioning that the improved knowledge on 
Cabibbo-suppressed decays of singly charmed baryons may have other 
important implications on the understanding of the $\xic$ lifetime. 
As shown in \cite{Guberina:2002fz}, it is possible to obtain a 
model-independent prediction of this lifetime once a reliable 
estimate of the decay rate of inclusive Cabibbo-suppressed decays 
of $\Lambda_{c}^{+}$ is available. Therefore, a more systematic and 
detailed approach to the Cabibbo-suppressed decays of singly charmed 
baryons is called for from both the experimental side, as a way of 
reducing systematic errors, and the theoretical side, as a way of 
obtaining model-independent predictions of the $\xic$ lifetime.

With the calculated level of the contribution of Cabibbo-suppressed 
$\omc$ decays, it is clear that this form of the systematic error 
in the determination of the $\xic$ lifetime cannot provide an 
explanation of the present disaccord between theory and experiment. 
To achieve agreement, a new layer of theoretical analysis will have 
to be uncovered and new experimental data will have to be compiled.   


\paragraph*{Acknowledgements} This work was initiated in discussions 
with our experimental collegues R. Kass (at the time from CLEO) and 
V. Eiges (BELLE). We would also like to thank J. Trampeti\'c for  
useful discussions. This work is supported by the Ministry of  
Science, Education and Sport of the Republic of Croatia under the 
contract No. 0098002.

\bibliography{xicbib}

\end{document}